\documentclass[prb,twocolumn,floatfix,showpacs,superscriptaddress]{revtex4-1}
\usepackage{graphics}
\usepackage{epsfig}
\usepackage{times}
\usepackage{bm}
\usepackage{braket}
\usepackage{color}

\renewcommand{\i}{{\rm i}}

\usepackage{amsmath}	% required for `\align' (yatex added)

\begin{document}
%\draft
\title{Ground state and low-energy excitations of the Kitaev-Heisenberg ladder}
\author{Cli\`o Efthimia Agrapidis}
\affiliation{Institute for Theoretical Solid State Physics, IFW Dresden, 01069 Dresden, Germany}
\author{Jeroen van den Brink}
\affiliation{Institute for Theoretical Solid State Physics, IFW Dresden, 01069 Dresden, Germany}
\affiliation{Department of Physics, Technical University Dresden, 01069 Dresden, Germany}
\affiliation{Department of Physics, Washington University, St. Louis, MO 63130, USA}
\author{Satoshi Nishimoto}
 \affiliation{Institute for Theoretical Solid State Physics, IFW Dresden, 01069 Dresden, Germany}
\affiliation{Department of Physics, Technical University Dresden, 01069 Dresden, Germany}

\date{\today}

\begin{abstract}
We study the ground state and low-lying excited states of the Kitaev-Heisenberg model on a ladder geometry using the density matrix renormalization group and Lanczos exact diagonalization methods. The Kitaev and Heisenberg interactions are parametrized as $K=\sin\phi$ and $J=\cos\phi$ with an angle parameter $\phi$. Based on the results for several types of order parameters, excitation gaps, and entanglement spectra, the $\phi$-dependent ground-state phase diagram is determined. Remarkably, the phase diagram is quite similar to that of the Kitaev-Heisenberg model on a honeycomb lattice, exhibiting the same long-range ordered states, namely rung-singlet (analog to N\'eel in 3D), zigzag, ferromagnetic, and stripy; and the presence of Kitaev spin liquids around the exactly solvable Kitaev points $\phi=\pm\pi/2$. We also calculate the expectation value of a plaquette operator corresponding to a $\pi$-flux state in order to establish how the Kitaev spin liquid extends away from the $\phi=\pm\pi/2$. Furthermore, we determine the dynamical spin structure factor and discuss the effect of the Kitaev interaction on the spin-triplet dispersion.
\end{abstract}

\maketitle

\section{Introduction}

Quantum spin liquids (QSLs) have been widely investigated in the last decades. In particular, the introduction of the Kitaev model and related spin liquid, so-called ``Kitaev spin liquid (KSL)'', in 2006 \cite{Kitaev06} has attracted great interest in the condensed matter community. The Kitaev model consists of Ising bond-direction dependent interactions (leading to an exchange frustration) on a honeycomb lattice. Amazingly, it is exactly solvable. Few years later, Jackeli and Khalliulin pointed out that a strong spin-orbit coupling in the $d^5$ transition metal compounds could bring the realization of this model~\cite{Jackeli09}. However, it is not simple to experimentally stabilise the KSL because even small amount of Heisenberg interaction present in real materials easily takes the system into a magnetically long range ordered (LRO) state. A model containing Kitaev and (nearest-neighbor) Heisenberg interactions is the so-called Kitaev-Heisenberg (KH) Hamiltonian. The interplay of these two interactions leads to non-integrable ground states and requires numerical methods in order to determine the magnetic properties of the low-energy states.

In the last decade, a growing number of investigations have been carried out on Kitaev materials~\cite{trebst17}. Generally, the candidates are classified broadly into two groups in terms of the geometries: Two-dimensional (2D) honeycomb materials such as $\alpha$-RuCl$_3$ \cite{Banerjee16, Do17, Hentrich18}, Na$_2$IrO$_3$, $\beta$-Li$_2$IrO$_3$ \cite{Singh12, Choi12, Takayama15, HwanChun15} and three-dimensional (3D) hyperhoneycomb materials such as $\alpha$- and $\gamma$-Li$_2$IrO$_3$ \cite{Katukuri16, Huang18}. Nonetheless, all these materials exhibit magnetic ordering at low temperature under normal pressures~\cite{Singh10, Johnson15}. Mostly, further interactions beyond the pure KH model, such as long range Heisenberg and off-diagonal exchange interactions, seem to play a crucial role in the magnetic properties. Recently, the possibility of a pressure and/or field induced spin liquid state has been also intensively studied. For this reason, testing a wide variety of internal and external parameters on the Kitaev materials has been a subject of active research. To evaluate the effect of such parameters correctly, a detailed understanding of the pure KH model is becoming more and more important in the context of QSL research.

\begin{figure}
  \includegraphics[width=0.9\columnwidth]{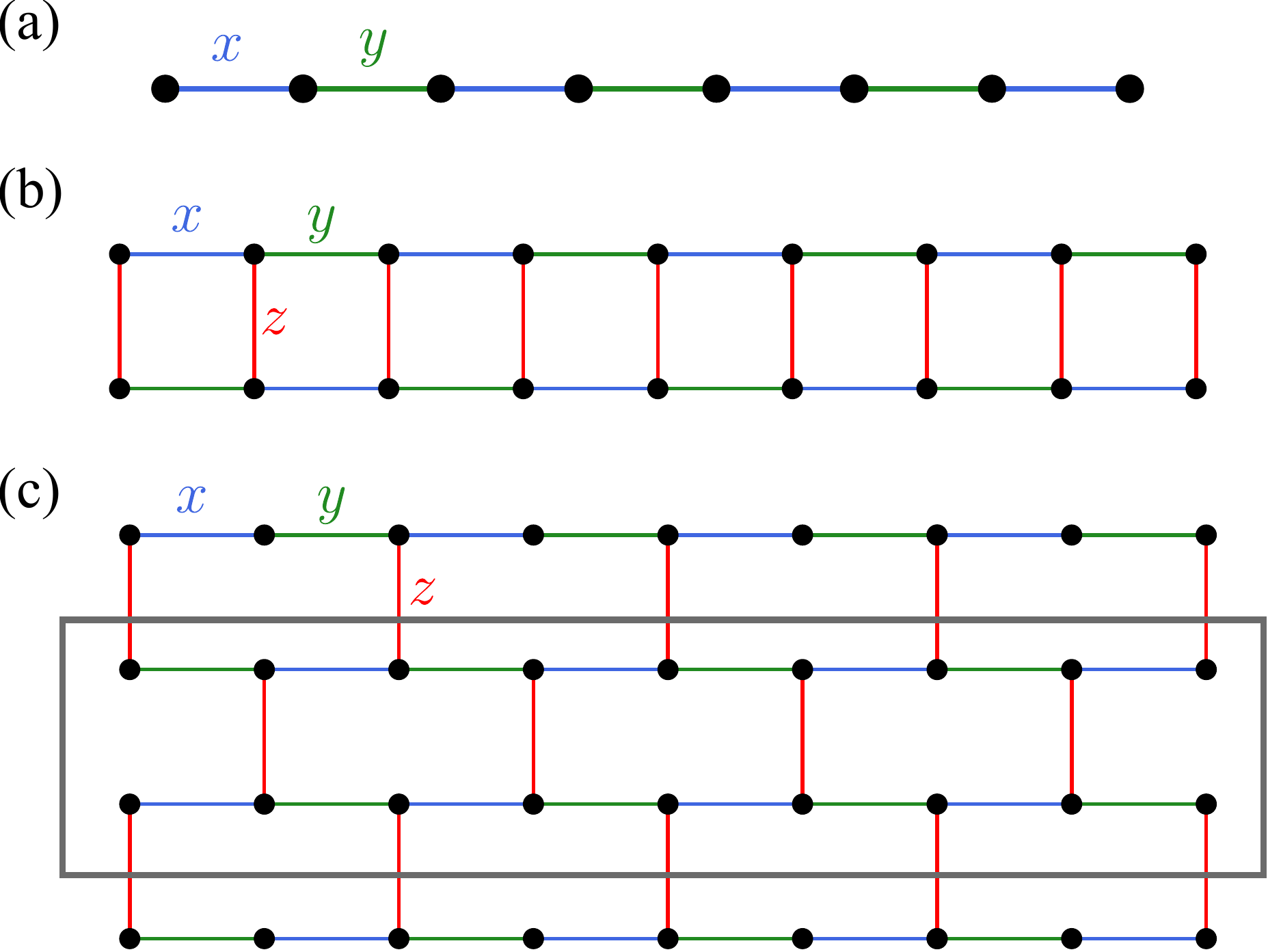}
  \caption{(a) Lattice structure of the KH chain. (b) Lattice structure of the KH ladder studied in this paper. (c) Structure of the KH model on a brickwall lattice, which is geometrically equivalent to the honeycomb lattice. The grey rectangle shows a cutout that makes the mapping to the ladder represented in (b) possible. The indices $x$, $y$ and $z$ indicate the three different bonds: $x$-bond, $y$-bond, and $z$-bond, respectively.
  }
  \label{lattice}
\end{figure}

Though the original KSL was introduced on the honeycomb lattice, it is known that the Kitaev interaction on any 3-coordinated lattice gives rise to non-trivial properties: In this sense, while a one-dimensional (1D) KH chain represented in Fig.~\ref{lattice}(a) cannot possess a KSL state, the KH model on a ladder (we simply refer to it as the KH ladder hereafter) in Fig.~\ref{lattice}(b) already meets the geometrical requirement. The KH ladder can be also extracted from a brickwall lattice [Fig.~\ref{lattice}(c) ], which is geometrically equivalent to the honeycomb lattice: Cutting along the grey line and folding the cut $z$-bonds toward the center we recover the KH ladder in Fig. \ref{lattice}(b). Therefore, it is expected that one can gain insight about the basic properties of the honeycomb-lattice KH model from the KH ladder. In fact, we previously found a certain similarity in magnetic ordering even between the 1D KH model and the honeycomb-lattice KH model~\cite{Agrapidis18}. Since the interplay of Kitaev and Heisenberg interactions in 2D or 3D geometries may pose serious challenges to the available numerical methods, it is a good strategy to consider the ladder system next. We can make use of he density-matrix renormalization group (DMRG) technique to study quasi-1D systems with great accuracy~\cite{white92}. Moreover, the ground-state properties and phase diagram of the coupled KH chains are yet to be extensively discussed~\cite{Metavitsiadis18}.

Motivated by this situation, we study the KH ladder using the DMRG method in this paper. We obtain the ground-state phase diagram to be composed of four magnetically ordered phases, namely rung-singlet, stripy, ferromagnetic (FM)-$xy$, zigzag; and two liquid phases, namely antiferromagnetic (AFM) KSL and FM KSL, depending on the ratio between Kitaev and Heisenberg interactions. To determine the phase boundaries, we compute several order parameters, excitation gap, and entanglement spectra. Strikingly, the phase diagram of the KH ladder is very similar to that of the honeycomb-lattice KH model. We then proceed at analyzing the low-lying excitations of the KH ladder by calculating the dynamical spin structure factor with using the Lanczos exact diagonalization (ED). It is interesting that most of the spectral features can be explained by considering those of the 1D KH model~\cite{Agrapidis18}.

The paper is organized as follows: Our Hamiltonian of the KH ladder is explained and the applied numerical methods are described in Sec. II. In Sec. III we present the four kinds of LRO magnetic state that are present depending on the ratio between Kitaev and Heisenberg interactions. In Sec. IV we discuss the properties of KSL states expanded around the large limit of Kitaev interaction. In Sec. V the ground states are summarized as a phase diagram as a function of the ratio between Kitaev and Heisenberg interactions. We also compare the ground-state phase diagram with those of 1D KH model and 2D honeycomb-lattice KH model, and discuss the similarity and dissimilarity among them. Sec. VI explains the fundamental features of dynamical spin structure factor in each the phase. Finally we conclude in Sec. VII.

\section{Model and Method}

\subsection{Model}

We study the KH ladder as represented in Fig.~\ref{lattice}(b). The Hamiltonian is described by
\begin{align}
  \mathcal{H}&= K \sum_{i=1}^{L/2}( S_{2i-1,1}^x S^x_{2i,1} + S^y_{2i,1}S^y_{2i+1,1}) + J \sum_{i=1}^L \vec S_{i,1} \cdot \vec S_{i+1,1}\nonumber \\
  & + K\sum_{i=1}^{L/2}(S^x_{2i,2}S^x_{2i+1,2}+S^y_{2i-1,2}S^y_{2i,2}) + J \sum_{i=1}^L \vec S_{i,2} \cdot \vec S_{i+1,2} \nonumber \\
  & + K \sum_{i=1}^{L} S^z_{i,1} S^z_{i,2} + J\sum_{i=1}^L \vec S_{i,1} \cdot \vec S_{i,2},
  \label{ham}
\end{align}
where $S_{i,j}^\alpha$ is the $\alpha$-component of spin-$\frac{1}{2}$ operator $\vec S_{i,j}$ at rung $i$ and leg $j$ ($=1,2$), $L$ is the system length, and $K$ and $J$ are the Kitaev and Heisenberg interactions, respectively. The first two lines denote the intra-leg interactions and the last line denotes the inter-leg interactions: Each leg has a period of two lattice spacing and there are three kinds of bond-dependent interactions. As shown in Fig.~\ref{lattice}(c) one finds that the KH ladder \eqref{ham} is a system cut out of the KH model on a Brickwall lattice. Since the Brickwall-lattice KH model is obtained by deforming the honeycomb-lattice KH model, the KH ladder is geometrically equivalent to the honeycomb-lattice KH model. Note that the KH ladder has a strong cluster anisotropy, i.e., the periodicity along the $z$ bond is short. Nonetheless, the LRO states observed in the honeycomb-lattice KH model also have a short periodicity in the bond direction and all of them can be reproduced in the KH ladder as shown below. In this paper, to compare magnetic properties of the KH ladder to those of the honeycomb-lattice KH model, we focus on the case of equal Kitaev and Heisenberg terms on the three bonds. For convenience, we introduce an angle parameter $\phi$, setting $J=\cos\phi$ and $K=\sin\phi$.

The Hamiltonian \eqref{ham} can be rewritten as
\begin{align}
  \mathcal{H}_{\rm leg}& = \frac{2J+K}{4} \sum_{j=1}^2 \sum_{i=1}^L (S^+_{i,j}S^-_{i+1,j}+S^-_{i,j}S^+_{i+1,j}) \nonumber \\
  & + \frac {K}{4} \sum_{j=1}^2 \sum_{i=1}^L (-1)^{(i+j)} (S^+_{i,j}S^+_{i+1,j} + S^-_{i,j}S^-_{i+1,j} )\nonumber \\
  & +J \sum_{j=1}^2 \sum_{i=1}^L S^z_{i,j} S^z_{i+1,j},
  \label{hamleg}
\end{align}
for the intra-leg couplings, and
\begin{align}
  \mathcal{H}_{\rm rung}& = \frac{J}{2} \sum_{i=1}^L (S^+_{i,1} S^-_{i,2} + S^-_{i,1}S^+_{i,2}) + (J+K) \sum_{i=1}^L S^z_{i,1}S^z_{i,2}
  \label{hamrung}
\end{align}
for the inter-leg, i.e., rung, couplings. We can easily notice that all the nearest-neighbor bonds have a XXZ-type interactions, and the sign-alternating double-spin-flip fluctuations acts only along the leg direction.

\subsection{Method}

We employ the DMRG method to investigate the ground-state properties of our model \eqref{ham}. We study finite-size systems with length up to $L\times 2 = 160\times 2$ with keeping up to $m=4000$ density-matrix eigenstates in the renormalization procedure. In this way, the truncation error, i.e. the discarded weight, is $\sim10^{-11}$. The calculated quantities are extrapolated to the limit $m \to \infty$ if needed. This allows us to perform very accurate finite-size scaling analysis. We use open and periodic boundary conditions depending on the quantity we consider. To identify the ground state for the given angle parameter $\phi$, we compute several order parameters, spin gap, plaquette operator, dynamical spin structure factor and entanglement spectra. When we calculate the order parameter under  open boundary conditions, the LRO state is observed as a state with a broken translational or spin symmetry. There are in fact several degenerate ground states; one configuration of the degenerate states is selected as the ground state by the initial condition of the DMRG calculation.

For the dynamical calculations, we use the Lanczos ED method. To examine the low-energy excitations for each phase, we calculate the dynamical spin structure factor, defined as
\begin{eqnarray}
\nonumber
S_\gamma(q,\omega) &=& \frac{1}{\pi}{\rm Im} \langle \psi_0 | (S^\gamma_q)^\dagger \frac{1}{\hat{H}+\omega-E_0-{\rm i}\eta} S^\gamma_q | \psi_0 \rangle\\
&=& \sum_\nu |\langle \psi_\nu |S^\gamma_q| \psi_0 \rangle|^2 \delta(\omega-E_\nu+E_0),
\label{spec}
\end{eqnarray}
where $\gamma$ is $z$ or $-(+)$, $| \psi_\nu \rangle$ and $E_\nu$ are the $\nu$-th eingenstate and the eigenenergy  of the system, respectively ($\nu=0$ corresponds to the ground state). Under periodic boundary conditions, the spin operators $S^\gamma_q$ can be precisely defined by
\begin{equation}
S^\gamma_q = \sqrt{\frac{2}{L}}\sum_iS^\gamma_{i,j}\exp(iqr_{i,j})
\label{sq}
\end{equation}
where $r_{i,j}$ is the position of site $(i,j)$. The sum runs over either $i$ even or $i$ odd sites with fixing $j=1$ or $2$. They provide the same results. The momentum is taken as $q=\frac{4\pi}{L}n$ ($n=0,\pm1,\dots, \pm\frac{L}{4}$) since the lattice unit cell includes four sites and the number of unit cells is $\frac{L}{2}$ in a system with $L \times 2$ sites. We calculate both spectral functions $S_\pm(q,\omega)$ and $S_z(q,\omega)$ as they are different due to broken SU(2) symmetry except at $\phi=0$ and $\pi$. We study ladders with $L \times 2=12 \times 2$, namely, 6 unit cells, by the Lanczos ED method. As shown below, our model \eqref{ham} contains only commensurate phases with unit cell containing one, two, or four sites. Therefore, a quantitative discussion for the low-lying excitations is possible even within the $12 \times 2$ ladder.

\section{Ordered phases}

In this section, we present the DMRG results for LRO phases in the ground state. We find four kinds of ordering, namely: stripy, rung-singlet, zigzag, and FM-$xy$ phases. The rung-singlet state is not magnetically ordered but the system is in a unique state with dimer ordering, namely, the dimer-dimer correlation is long ranged. The names of the ordered phases follow Ref.~\onlinecite{Choi12}. In the LRO states, except for the rung-singlet state, the translational or spin rotation symmetry is broken in a finite system due to Friedel oscillations under open boundary conditions, so that the ordered state can be directly observed with a local quantity by extracting one of the degenerate states. Generally, the Friedel oscillations in the center of the system decay as a function of the system length. If the amplitude at the center of the system persists for arbitrary system lengths, it corresponds to a long range ordering.

\subsection{Stripy phase $(1.57\pi < \phi < 1.7\pi)$}

\begin{figure}
  \includegraphics[width=0.9\columnwidth]{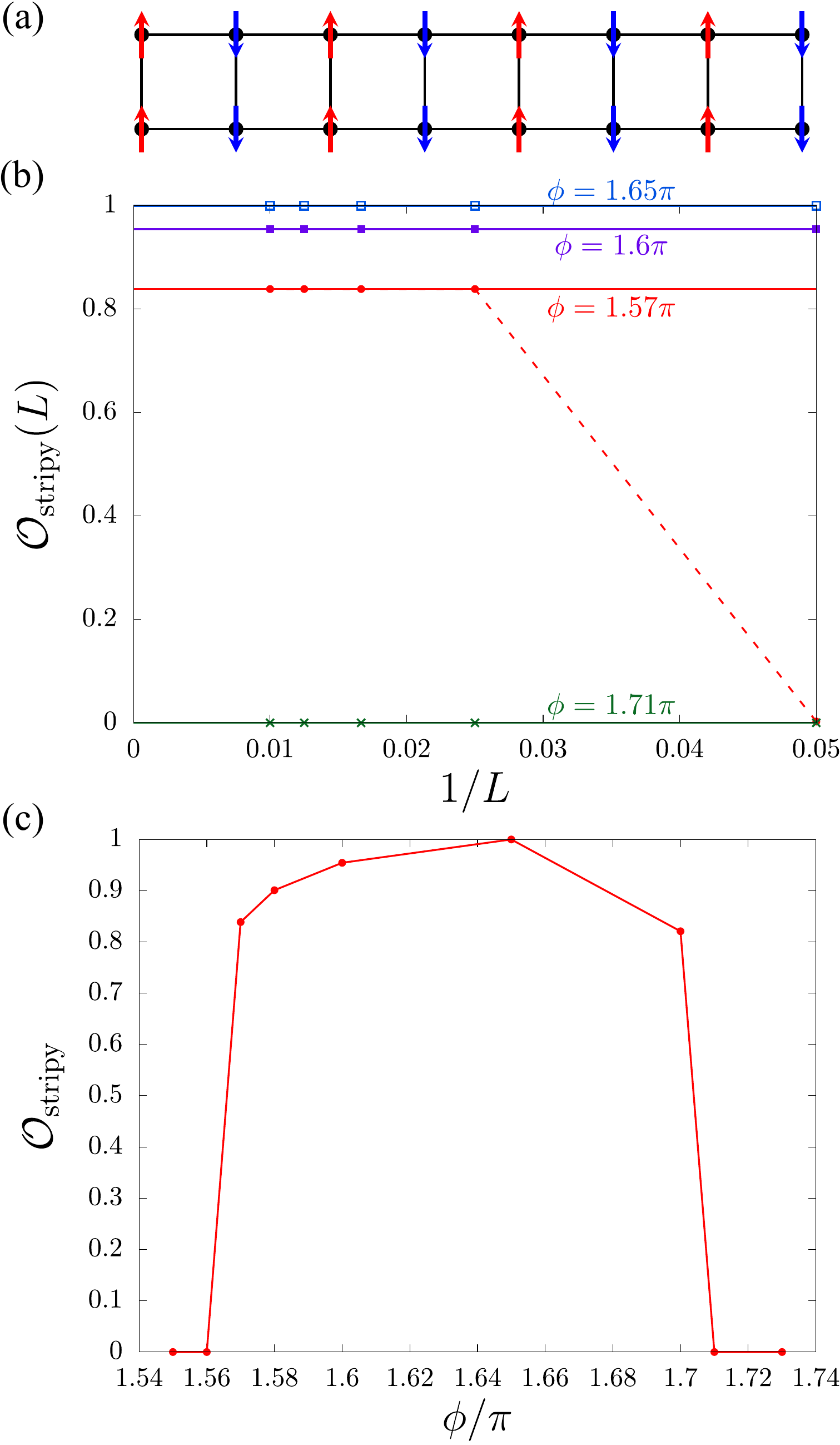}
  \caption{(a) Schematic spin configuration of the stripy state. (b) Finite size scaling of the stripy order parameter for several values of $\phi$. The dotted line helps the eye follow the data points, solid lines represent the linear fitting. (c) Extrapolated stripy order parameter as a function of $\phi/\pi$.}
  \label{stripy}
\end{figure}

Let us start with the stripy state. In the region of $\frac{3}{2}\pi \lesssim \phi < \frac{7}{4}\pi$, since $J>0$ and $J+K<0$, the leg and rung interactions are AFM and FM, respectively. Thus, we naively expect the coupled chains to order in what we call the stripy state, as depicted in Fig.~\ref{stripy}(a). Getting back to the original Brickwall lattice, the alignments of up spins and down spins appear alternately with running along the leg. This state can be analytically proven at $\phi=\tan^{-1}(-2)\approx1.65\pi$, where our model \eqref{ham} is exactly solvable: The rung Hamiltonian \eqref{hamrung} leads simply to isotropic FM couplings due to $J+K=-J$; whereas, the leg Hamiltonian \eqref{hamleg} is reduced to a sum of double-spin-flip ($S^+S^++S^-S^-$) and Ising ($S^zS^z$) parts because the exchange ($S^+S^-+S^-S^+$) term disappears due to $2J+K=0$. The total energy of our system \eqref{ham} is minimized by taking the wave function as
\begin{align}
|\Psi_0\rangle = \frac{1}{\sqrt{2}}\left[\prod_{i=1}^{L/2} S^-_{2i,1}S^-_{2i,2}|\Uparrow\rangle + \prod_{i=1}^{L/2} S^+_{2i,1}S^+_{2i,2}|\Downarrow\rangle\right],
\label{wf_stripy}
\end{align}
where $|\Uparrow\rangle$ and $|\Downarrow\rangle$ denote configurations including only up and down spins, respectively. Note that all the spins are aligned along the $z$-direction. When the Hamiltonian \eqref{ham} is applied to this wave function \eqref{wf_stripy}, only the Ising terms provide nonzero components. Thus, no quantum fluctuations exist and the system is in a perfect stripy state described by Eq.~\eqref{wf_stripy}. The ground state energy is $E_0/L=\frac{3}{4}J$.

It is still a nontrivial question how the wave function~\eqref{wf_stripy} is modified with moving away from $\phi \approx 1.65\pi$. To study it numerically, we introduce an order parameter defined by
\begin{align}
\mathcal O_{\mathrm{stripy}} (L)&= \frac{1}{2}\left(| \langle S^z_{(L/2,1)}\rangle - \langle S^z_{(L/2+1,1)}\rangle \right.  \nonumber\\
&\left. + \langle S^z_{(L/2,2)}\rangle -\langle S^z_{L/2+1,2)}\rangle|\right) \\
\mathcal O_{\mathrm{stripy}}&=\lim_{L\to\infty} \mathcal O_{\mathrm{stripy}} (L)
\end{align}
In Fig.~\ref{stripy}(b), we show finite-size scaling analysis of $\mathcal O_{\mathrm{stripy}}$. We see how finite-size scaling is of fundamental importance in this system: the dotted line shows the jump in the order parameter between the smallest ($20\times2$) and the second smallest ($40\times2$) systems. The finite size scaling is then performed with discarding the first point, where the system size is too small to stabilize the ordering. This explicitly indicates the existence of ``critical length'' for stabilizing a long range order. Example of this kind of behavior are seen also for order parameters of the other ordered states. The $L \to \infty$ extrapolated value of the stripy order parameter is plotted in Fig.~\ref{stripy}(c). The validity of the exact wave function \eqref{wf_stripy} is confirmed by $\mathcal O_{\mathrm{stripy}}=1$ at $\phi\approx1.65\pi$. Even away from $\phi\approx1.65\pi$, $\mathcal O_{\mathrm{stripy}}$ keeps relatively large value ($\sim 1$) and drops down to zero at the both edges, $\phi\approx1.57\pi$ and $\phi=1.7\pi$. It means that the transitions at both phase boundaries are of the first order.

In our previous paper~\cite{Agrapidis18} we found that the 1D KH model exhibits a Ne\'el-$z$ state, i.e., Ne\'el ordering with spins parallel or antiparallel to the $z$-axis, for $1.65\pi \lesssim \phi < 2\pi$. In this sense the stripy state oh the KH ladder may be also interpreted as two Ne\'el-$z$ chains coupled by FM interaction.

\subsection{Zigzag phase $(0.53\pi \leq \phi <0.8\pi)$}

\begin{figure}
  \includegraphics[width=0.9\columnwidth]{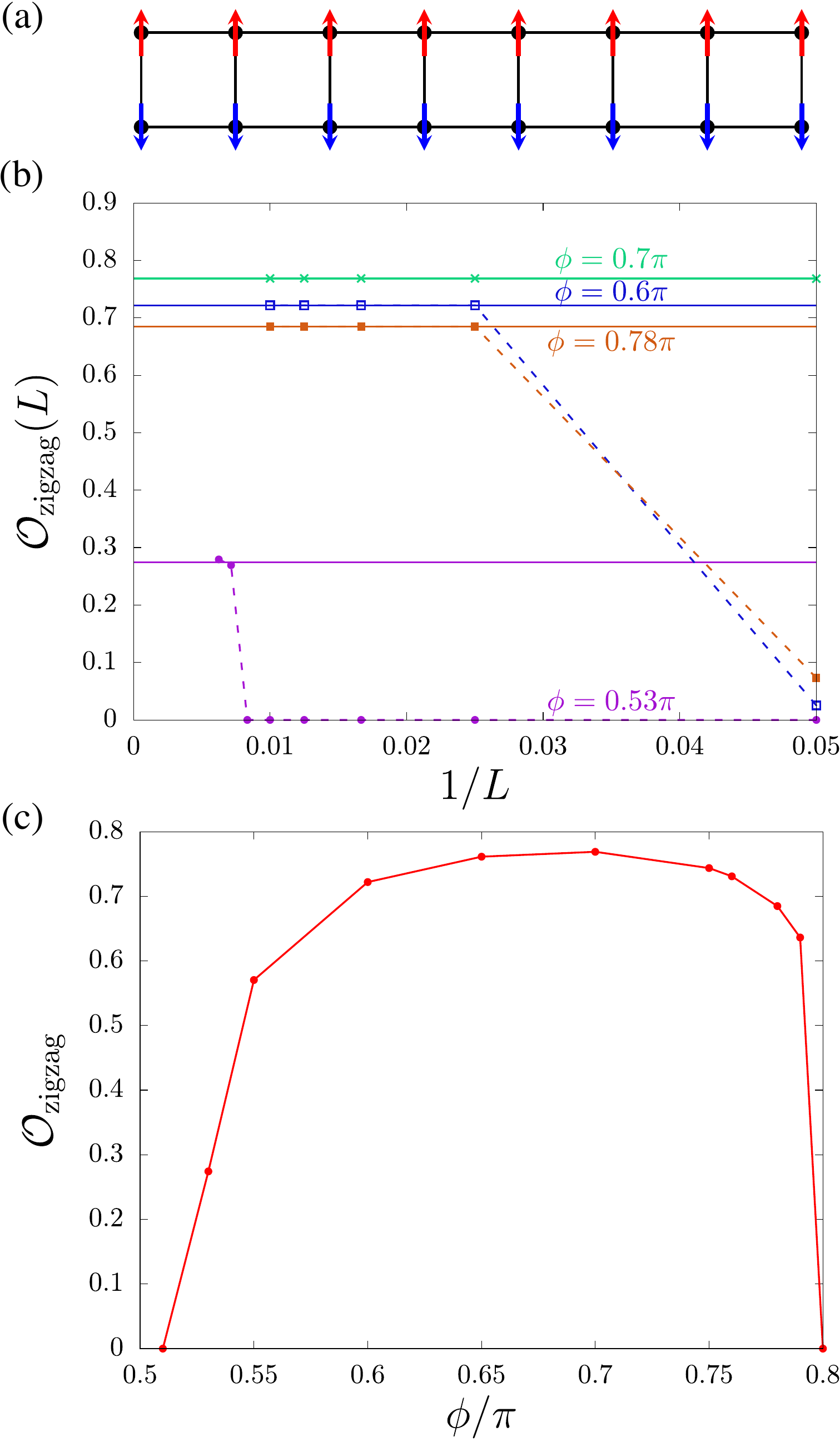}
  \caption{(a) Schematic spin configuration of the zigzag state. (b) Finite size scaling of the zigzag order parameter for several values of $\phi$. Dotted lines help the eye follow the data points, solid lines represent the linear fitting. Note that for $\phi=0.53\pi$ the order parameters is finite only for $L\geq140$. (c) Extrapolated zigzag order parameter as a function of $\phi/\pi$.}
  \label{zigzag}
\end{figure}

In the region of $\frac{1}{2}\pi \lesssim \phi < \frac{3}{4}\pi$, since $J<0$ and $J+K>0$, the leg and rung interactions are FM and AFM, respectively. Hence, an ordered state as in Fig.~\ref{zigzag}(a) is expected. We call it {\it zigzag} state by following the name of the corresponding state in the honeycomb-lattice KH model~\cite{Choi12} (our {\it straight} leg corresponds to a zigzag line in the honeycomb lattice). Through a similar analysis of the exact wave function \eqref{wf_stripy} at $\phi\approx1.65\pi$, we could assume the wave function at $\phi\approx0.65\pi$ to be
\begin{align}
|\Psi_0\rangle \approx \frac{1}{\sqrt{2}}\left[\prod_{i=1}^L S^-_{i,1}|\Uparrow\rangle + \prod_{i=1}^L S^+_{i,1}|\Downarrow\rangle\right].
\label{wf_zigzag}
\end{align}
However, unlike in the case of $\phi\approx1.65\pi$, this {\it classical} configuration is just a good approximation for the wave function at $\phi\approx0.65\pi$ but not an exact one because quantum fluctuations are involved from the intra-leg double-spin-flip and rung exchange processes.

We define the following order parameter to see the instability of zigzag ordering:
\begin{align}
	\mathcal O_{\mathrm{zigzag}} (L)&= \frac{1}{2}\left(| \langle S^z_{(L/2,1)}\rangle + \langle S^z_{(L/2+1,1)}\rangle \right.\nonumber \\
	&\left.- \langle S^z_{(L/2,2)}\rangle -\langle S^z_{L/2+1,2)}\rangle| \right) \\
	\mathcal O_{\mathrm{zigzag}}&=\lim_{L\to\infty} \mathcal O_{\mathrm{zigzag}} (L)
\end{align}
Fig.~\ref{zigzag}(b)(c) show the finite-size scaling analysis of $\mathcal O_{\mathrm{zigzag}} (L)$ for several values of $\phi$. At the lower boundary with the AFM KSL phase ($\phi=0.53\pi$), the long range order settles only at large system sizes $L\geq140$: This can be interpreted as some kind of ''fragility`` of the zigzag ordering close to the AFM KSL. Moreover, it underlines the importance of studying large enough ladders using the DMRG method for this system. In Fig.~\ref{zigzag}(c) we plot the extrapolated values of $\mathcal O_{\mathrm{zigzag}}$ in the thermodynamic limit. We can see that $\mathcal O_{\mathrm{zigzag}}$ keeps $\sim 0.7-0.8$ in most of the zigzag phase and Eq.~\eqref{wf_zigzag} gives a good approximation for this zigzag state. Around the lower phase boundary ($\phi\sim0.53\pi$), $\mathcal O_{\mathrm{zigzag}}$ approaches rather continuously to zero with approaching the phase boundary, suggesting a second order of continuous transition; whereas at the upper phase boundary ($\phi\sim0.8\pi$), $\mathcal O_{\mathrm{zigzag}}$ drops down to $0$, suggesting a first-order transition.

For $\phi<0.75\pi$, the leading interaction on the rungs is AFM since $J+K$ is positive in Eq.~\eqref{hamrung}. Therefore, the zigzag state may be simply interpreted as antiferromagnetically coupled FM chains (``FM-$z$ state''~\cite{Agrapidis18}) obtained in the 1D KH model at $0.65\pi<\phi<\pi$. However, the lower bound of the zigzag phase is significantly more extended (down to $\phi=0.53\pi$) than the lower bound of the FM-$z$ state ($\phi=0.65\pi$) in the 1D KH model. At $0.5\pi<\phi<0.65\pi$ the 1D KH model is in a liquid state called ``spiral-$xy$ state''. Nevertheless, ferromagnetic fluctuations on the legs would be strong because of the negative $J$ in Eq.~\eqref{hamleg} and the zigzag ordering can be stabilized by the dominant AFM Ising term on the rungs due to $J+K>|J|$ in Eq.~\eqref{hamrung}. In other words, the FM alignment on each leg is just taken care of by the interchain AFM couplings. This may be related to the fragility of the zigzag order near the AFM KSL phase.

\subsection{Rung-singlet phase ($-0.3\pi \leq \phi \leq 0.48\pi$)}

\begin{figure}
  \includegraphics[width=0.9\columnwidth]{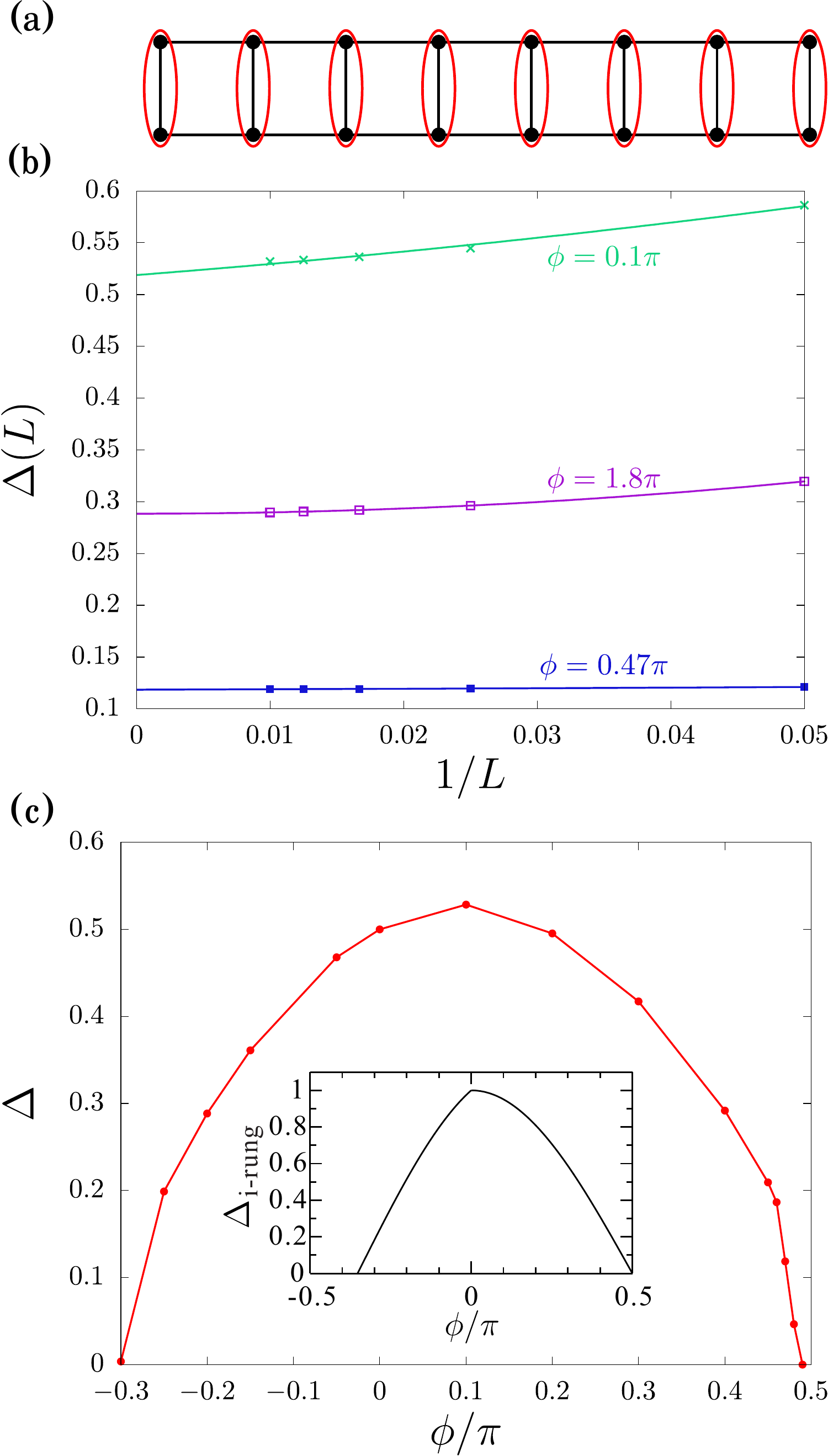}
  \caption{(a) Schematic spin configuration of the rung singlet state, where a red ellipse represents a spin singlet. (b) Finite size scaling of the gap for $\phi=-0.2\pi$, $0.1\pi$, and $0.47\pi$. (c) Extrapolated spin gap as a function of $\phi/\pi$. Inset: Spin gap of isolated rung as a function of $\phi$.}
  \label{rungS0}
\end{figure}

At $\phi=0$ ($J=1$, $K=0$), our system \eqref{ham} is a pure isotropic AFM Heisenberg ladder, known to be in a rung-singlet state with singlet-triplet excitation gap $\Delta=0.5037J$ (Ref.~\onlinecite{White94}). The schematic picture of rung-singlet state is given in Fig.~\ref{rungS0}(a). We compute the excitation gap to see how the perturbation introduced by the Kitaev term affects this state. Since the total $S^z$ is not a good quantum number except at $\phi=0$, the gap is simply defined as the energy difference between the ground state and first excited state:
\begin{equation}
 \Delta(L)=E_1(L)-E_0(L) \qquad \Delta=\lim_{L\to\infty} \Delta(L),
 \label{eq-gap}
\end{equation}
where $E_0$ is the ground state energy and $E_1$ is the first excited state energy. Fig.~\ref{rungS0}(b)(c) show the finite-size scaling analysis of $\Delta (L)$ for several values of $\phi$ and the $L\to\infty$ extrapolated value of $\Delta$ as a function of $\phi$, respectively. It is remarkable that the gap is clearly asymmetric about $\phi=0$, reaching its maximum at $\phi\sim0.1\pi$: this could be understood by noticing that both $K$ and $J$ are AFM in the region of $0<\phi<\frac{1}{2}$; while, $K$ and $J$ have different signs in the region of $-\frac{1}{2}\pi<\phi<0$. The gap closes gradually with approaching the boundary to the stripy phase at $\phi=-0.3\pi$ and to the AFM KSL phase at $\phi=0.48\pi$. Thus, they are both continuous transitions.

Let us provide a more comprehensive explanation about the asymmetry of the gap in respect to $\phi$. In an AFM Heisenberg ladder, it is known that the magnitude of the gap roughly scales with the AFM rung interaction. This also means that the spin-spin correlations are strongly screened. Therefore, a single dimer may be expected to be an effective model to reproduce the gap behavior. We then extract an isolated rung: $\frac{J}{2}(S_1^+S_2^-+S_1^-S_2^+)+(J+K)S_1^zS_2^z$ from our system \eqref{ham}. This two-site system can be easily diagonalized and the gap is obtained as $\Delta_{\rm i-rung}=\frac{2J+K}{2}$ for $\phi<0$ and $\Delta_{\rm i-rung}=J$ for $\phi>0$. In the inset of Fig.~\ref{rungS0}(c) the gap obtained for the isolated rung is plotted. The qualitative trend the of gap with $\phi$ is well described by the single dimer. Furthermore, the gap closing points at $\phi=\tan^{-1}(-2)\approx-0.35\pi$ and $\phi=\frac{1}{2}\pi$
are very close to those for the original KH ladder. It proves the strong screening of spin-spin correlations in the whole rung-singlet phase.

\subsection{Ferromagnetic phases}

\begin{figure}
  \includegraphics[width=0.9\columnwidth]{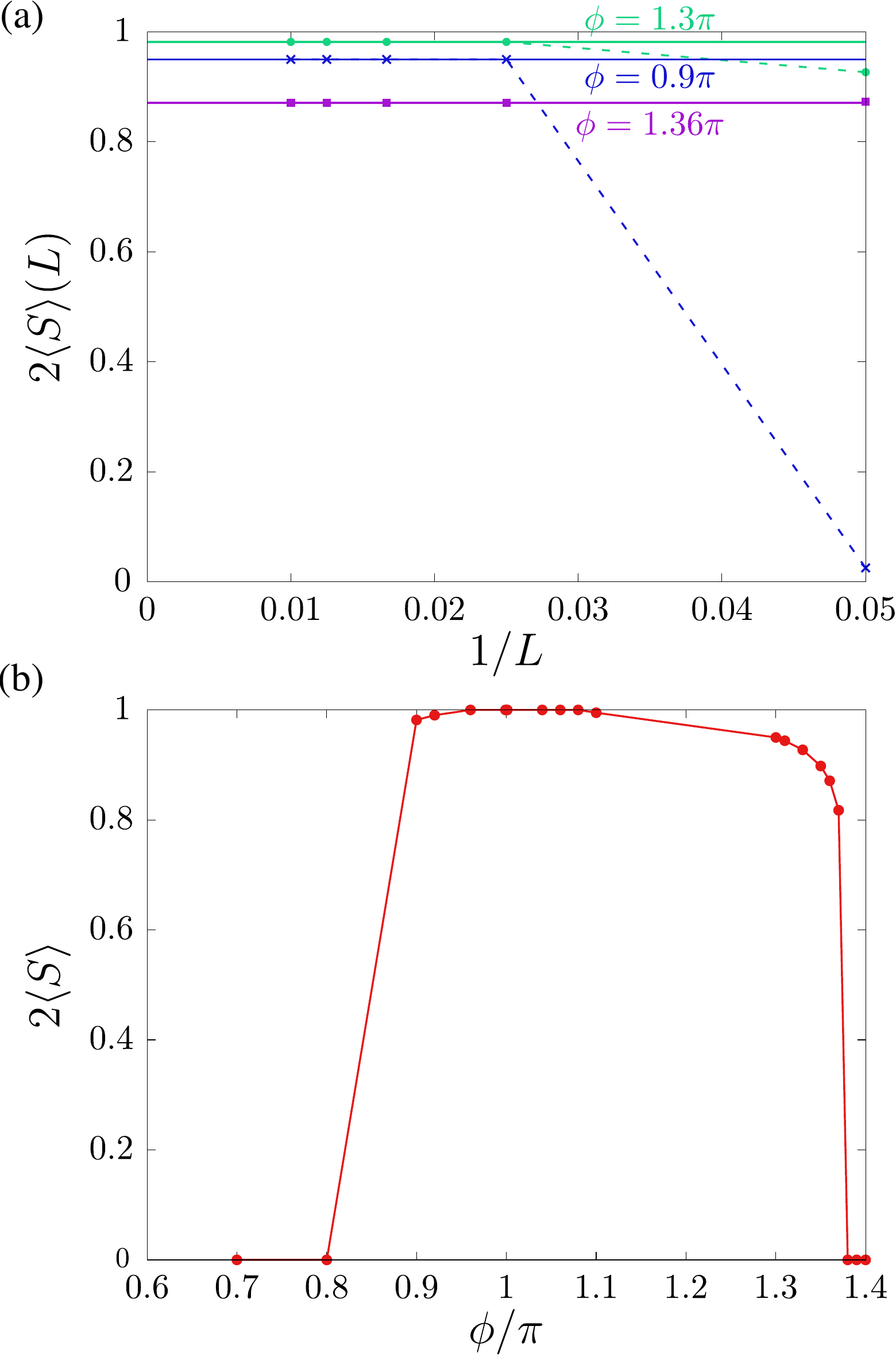}
  \caption{(a) Finite-size scaling of the local spin $\langle S \rangle$ for several values of $\phi$. Dotted lines help the eye follow the data points, solid lines represent the fitting. (b) Extrapolated values of $\langle S \rangle$ as a function of $\phi/\pi$.}
  \label{FM}
\end{figure}

At $\phi=\pi$ ($J=-1$, $K=0$), the system is in an SU(2) isotropic FM state. This state can be expressed as a sum of fully polarized spin configurations for all of the total $S^z$ sectors, namely, $S_{\mathrm{tot}}^z=\sum_i S^z_i=0,\pm1,\pm2,\cdots,\pm L$. On leaving $\phi=\pi$, they are lifted: For $|\pi-\phi|\ll 1$, the first order perturbation in the Hamiltonian \eqref{ham} is given by
\begin{align}
  \mathcal {H}'&=\frac{\pi-\phi}{4} \sum_{j=1}^2 \sum_{i=1}^L (S^+_{i,j}S^-_{i+1,j} + S^-_{i,j}S^+_{i+1,j}) \nonumber\\
  & + \frac{\pi-\phi}{4} \sum_{j=1}^2 \sum_{i=1}^L (-1)^{(i+j)} (S^+_{i,j}S^+_{i+1,j} + S^-_{i,j}S^-_{i+1,j} )\nonumber \\
  	             & + (\pi-\phi) \sum_{i=1}^L S^z_{i,1}S^z_{i,2},
                 \label{pham}
\end{align}
where the non-perturbative part is the simple FM Heisenberg ladder with $J=-1$ ($\forall$ nearest neighbor bonds). To gain the energy benefits by $\mathcal {H}'$, the total $S^z$ sectors in the wave function are restricted to $S_{\mathrm{tot}}^z=0,\pm2,\pm4,\cdots,\pm L$. Thus, near the vicinity of $\phi=\pi$ the ground state is approximately denoted by
\begin{equation}
 \ket \psi \approx \frac{1}{\sqrt {\mathcal N}} \sum_m \ket {\phi_m}
 \label{FM-wf}
\end{equation}
where $m$ runs over all the possible spin configurations $\ket {\phi_m}$ ($m=1\cdots\mathcal N$) with $S_{\mathrm{tot}}^z=0,\pm2,\pm4,\cdots,\pm L$, $\mathcal N$ is the number of the spin configurations, i.e.,  $\mathcal N =\sum_{n=0}^L {}_{2L} C _{2n}=\sum_{n=0}^L \frac{(2L)!}{(2n)! (2L-2n)!}$. As a result, the polarized direction is [110] in the spin space, namely, $\langle S_i^xS_j^x\rangle=\langle S_i^yS_j^y\rangle=\frac{1}{8}$ and $\langle S_i^zS_j^z\rangle=0$ ($\forall i,j$). We call it FM-$xy$ state. This breaking of the SU(2) symmetry is a consequence of the double-spin-flip term, which immediately suppressed the spin polarization along the $z$-axis.

To determine the range of the FM-$xy$ phase, we calculate the total spin per rung $S^{\rm tot}/(2L)$, defined by
\begin{align}
S^{\rm tot}(S^{\rm tot}+1)=\sum_{j,j'=1}^2 \sum_{i,i'=1}^L \vec{S}_{i,j}\cdot\vec{S}_{i',j'} %S^x_{i,j}S^x_{i',j'}+S^y_{i,j}S^y_{i',j'}+S^z_{i,j}S^z_{i',j'},
\label{FMparameter2a}
\end{align}
and the local spin
\begin{align}
\langle S \rangle=\sqrt{\langle S^x_{i,j} \rangle^2+\langle S^y_{i,j} \rangle^2+\langle S^z_{i,j} \rangle^2}
\label{FMparameter2b}
\end{align}
at the center of the system $i=\frac{L}{2}$. Note that we can directly detect the local moment in the FM state since the spin rotation symmetry is broken by using open boundary conditions. We have confirmed $\langle S \rangle=S^{\rm tot}/(2L)$ in the thermodynamic limit. Fig.~\ref{FM} shows $2\langle S \rangle$ as a function of $\phi$. At the isotropic SU(2) point ($\phi=\pi$), $2\langle S \rangle=S^{\rm tot}/L=1$. We find that $2\langle S \rangle$ decays very slowly from $1$ as the distance from $\phi=\pi$, and keeps $\sim1$ in the whole FM-$xy$ region $0.8\pi<\phi<1.37<\pi$. The robustness of the FM-$xy$ state is naively expected because both $J$ and $J+K$ are FM at $\frac{3}{4}\pi<\phi<\frac{3}{2}\pi$. Then, at both boundaries, to the zigzag state at $\phi=0.8\pi$ and to the spin liquid state $\phi=1.37\phi$, it sharply drops down to $0$, which indicates first order transitions.

\section{Spin liquid states}

We have determined the phase boundaries of LRO phases covering most of the $\phi$ range. In the remaining two narrow $\phi$ regions around the Kitaev points $\phi=\pm\frac{\pi}{2}$, we found no long range ordering, i.e., they are {\it spin liquid} states. To consider the similarity to the so-called KSL in the honeycomb KH model, we compute the expectation value of plaquette operator and the excitation gap.

\subsection{Plaquette operator}

\begin{figure}
  \includegraphics[width=0.9\columnwidth]{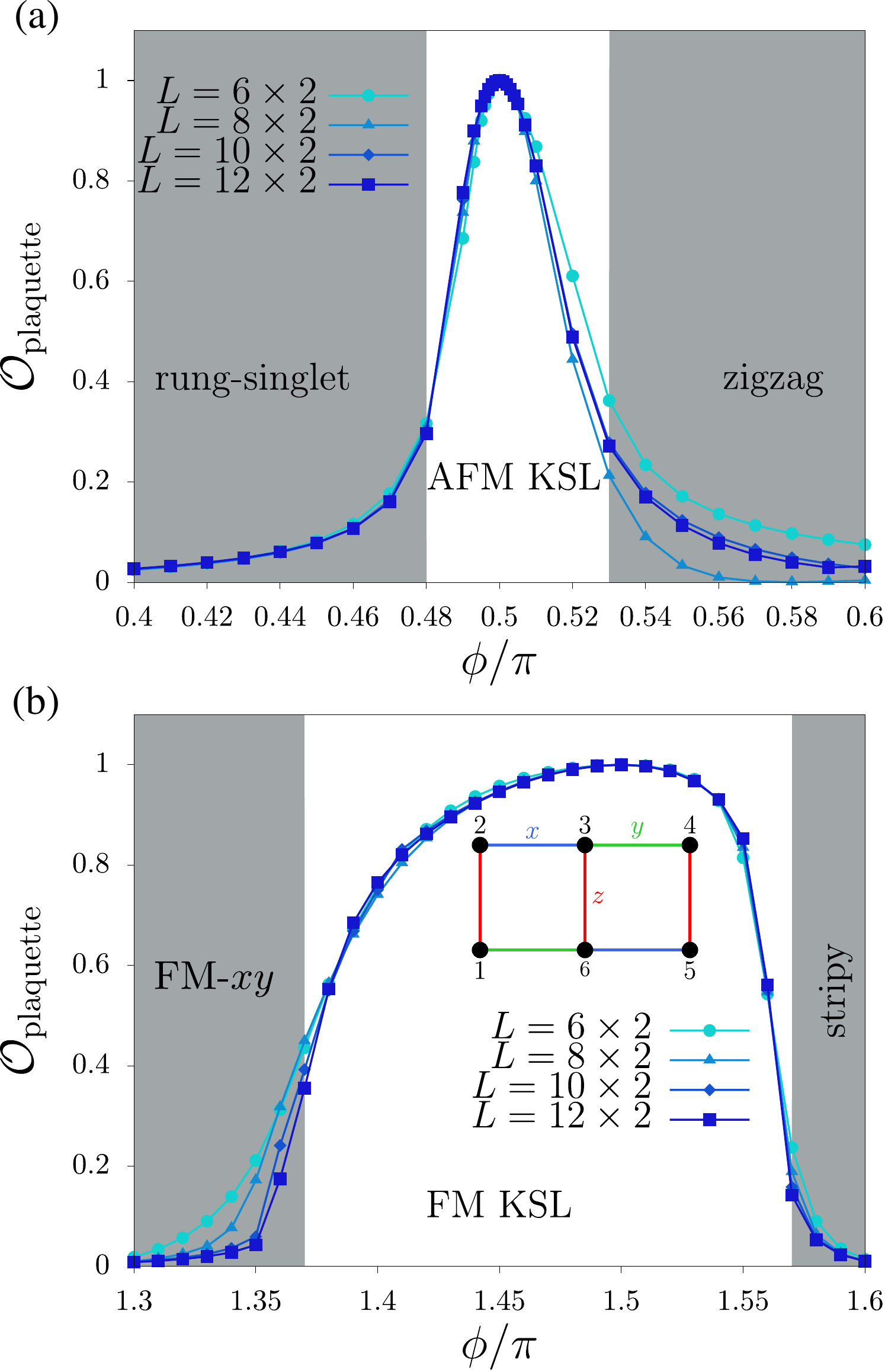}
  \caption{Expectation value of the plaquette operator around (a) the AFM KSL point $\phi=\pi/2$ and (b) the FM KSL point $\phi=3\pi/2$ for different system lengths. Shaded areas show the neighboring LRO phases. The insets in (a) and (b) show the considered 6-site plaquette corresponding to a hexagon in the honeycomb-lattice KH model.}
  \label{plaquette}
\end{figure}

It is known that the Kitaev model, e.g., on a hexagonal cluster and ladder, is in a $\pi$-flux state. This state is characterized by the expectation value of plaquette operator to be unity. We define the expectation value of the plaquette operator for our system \eqref{ham} as
\begin{equation}
 \mathcal{O}_{\mathrm{plaquette}}=\langle S_1^xS_2^yS_3^zS_4^xS_5^yS_6^z \rangle \label{plaquetteoperator}
\end{equation}
where the numbering of sites is indicated in the inset of Fig.~\ref{plaquette}. Note that this 6-site plaquette corresponds to a hexagon in the honeycomb-lattice KH model. In Fig.~\ref{plaquette} we show $\mathcal{O}_{\mathrm{plaquette}}$ calculated with ED for several ladder lengths under periodic boundary conditions. The finite-size effect seems to be negligible within the spin liquid phases. At both of the Kitaev points $\phi=\pm \pi/2$, $\mathcal{O}_{\mathrm{plaquette}}$ is 1 as expected. With moving away from $\phi=\pm \pi/2$, $\mathcal{O}_{\mathrm{plaquette}}$ decreases but keeps $\sim 1$ in finite regions. In the vicinities of the neighboring LRO phases, it decreases rapidly to $\sim 0$. This means that the ranges of spin liquid phases characterized by nonzero $\mathcal{O}_{\mathrm{plaquette}}$ are consistent to the phase boundaries with LRO state estimated by order parameters and spin moment. Interestingly, the region of FM KSL is a few times wider than that of AFM KSL. This is similar to the trend in the honeycomb KH model (see below). We have also confirmed that the spin-spin correlations are finite only between neighboring sites at the Kitaev points, as in the honeycomb-lattice KH model.

\subsection{Excitation gap}

\begin{figure}
  \includegraphics[width=0.9\columnwidth]{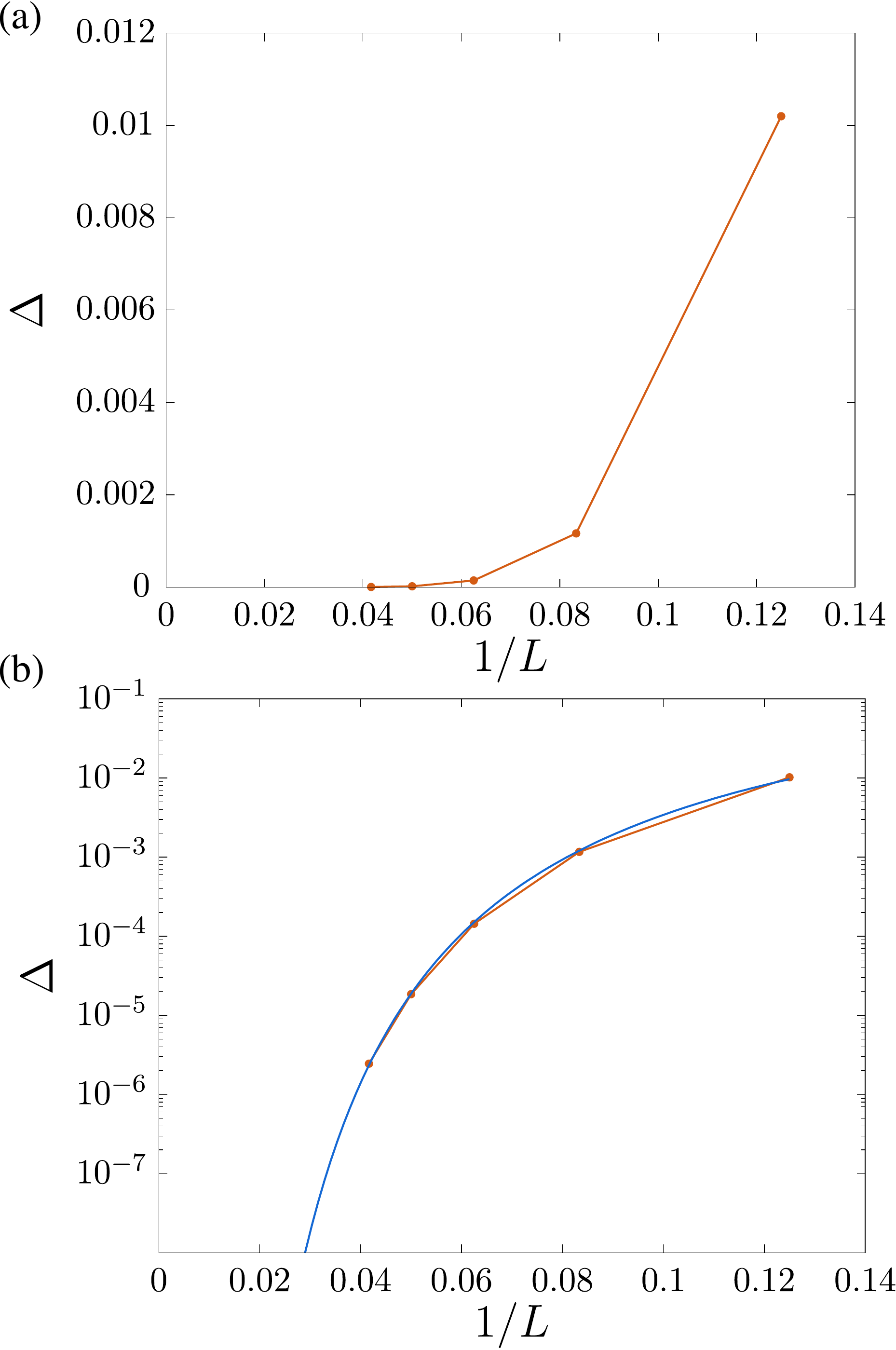}
  \caption{(a) Finite-size scaling of the excitation gap as a function of the inverse system length at the Kitaev points $\phi=\pm\pi/2$. (b) Semi-log plot of (a). The blue line is fitting function $\Delta=0.62\exp(-0.52L)$.}
  \label{gap}
\end{figure}

We compute the excitation gap using Eq. \eqref{eq-gap} at the two Kitaev points $\phi=\pm\pi/2$. The results are the same at both points. In Fig.~\ref{gap}(a) the excitation gap is plotted as a function of the inverse system size. It seems to indicate a vanishing gap in the thermodynamic limit. Actually, as shown in Fig.~\ref{gap}(b) an exponential decay is clearly seen by plotting it in a semi-log scale. Although this result might seem in opposition with the previous studies in Refs.~\onlinecite{Wu12, Feng07}, we can suggest at least that no gap exists between the ground state and the first excited state. Within our numerical analyses for the spin liquid state in the KH ladder we have found no difference from an isotropic KSL state in the honeycomb-lattice KH model. This should be further investigated in future studies.

\begin{figure*}
  \includegraphics[scale=0.55]{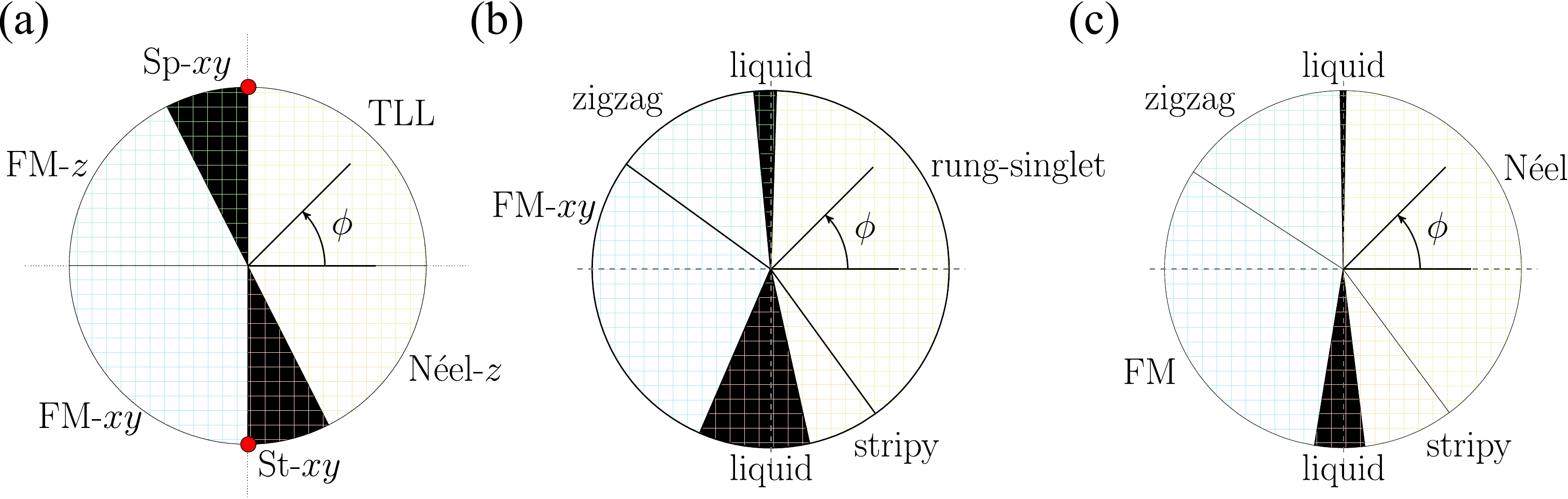}
  \caption{(a) $\phi$-dependent Phase diagram of the 1D KH model, where Sp-$xy$, St-$xy$, and TLL are abbreviations for ``spiral-$xy$'', ``staggered-$xy$'', and Tomonaga-Luttinger liquid, respectively. The details are explained in Ref.~\onlinecite{Agrapidis18}. (b) Phase diagram of the KH ladder, obtained in this paper. (c) Phase diagram of the honeycomb-lattice KH model~\cite{Chaloupka13}.}
  \label{phasediagram}
\end{figure*}

\begin{figure}
  \includegraphics[width=0.9\columnwidth]{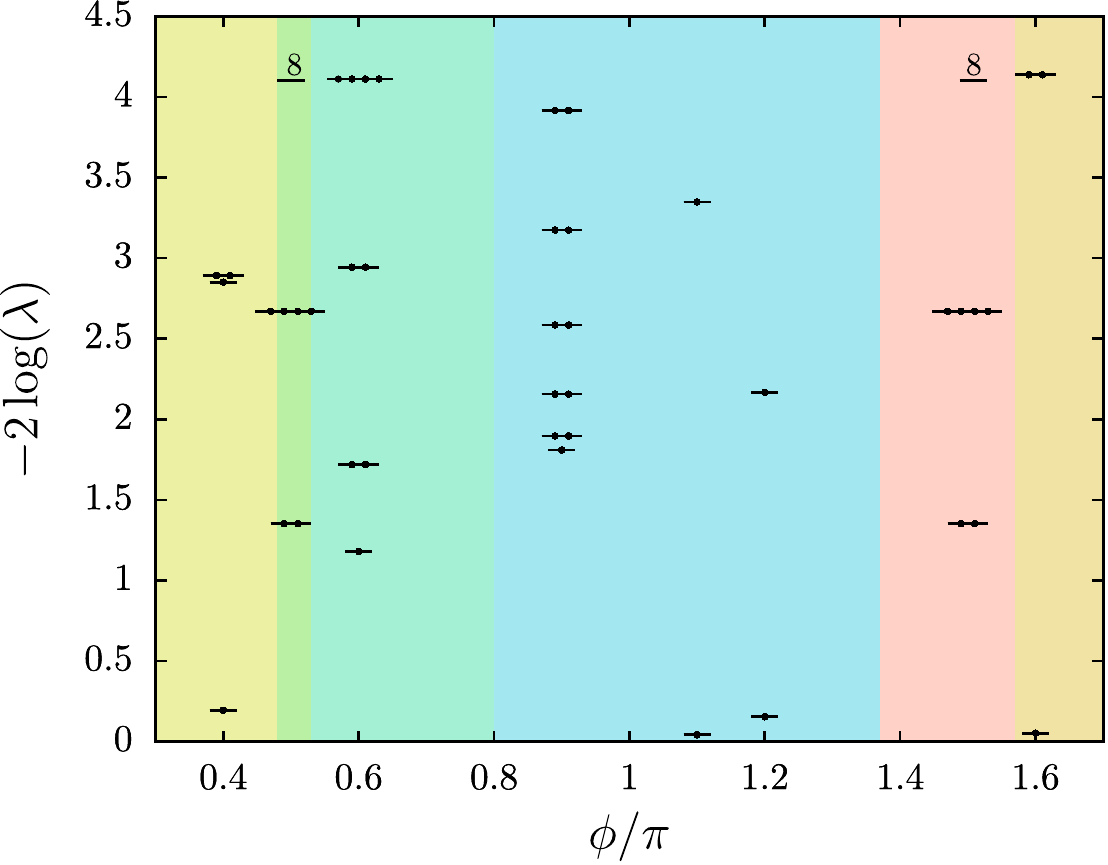}
  \caption{Entanglement spectra for representative $\phi$-points of the different phases in the ground-state phase diagram. The used system size is $L\times2=32\times2$ for the LRO states and $L\times2=24\times2$ for the two Kitaev points.}
  \label{ES}
\end{figure}

\section{Phase diagram}

Based on the above numerical results, we present the $\phi$-dependent ground-state phase diagram of the KH ladder in Fig.~\ref{phasediagram}(b). For comparison, we also show the ground-state phase diagrams of the KH model on a single chain~\cite{Agrapidis18} and on a honeycomb lattice~\cite{Chaloupka13} in Fig.~\ref{phasediagram}(a) and (c), respectively. In our previous paper~\cite{Agrapidis18}, we argued that the $\phi$-dependent phase diagram of the 1D KH model is similar to that of the 2D honeycomb-lattice KH model; all the LRO states of the honeycomb-lattice KH model can be interpreted in terms of the coupled KH chains. In this paper, surprisingly, we found that the phase diagram of just two coupled KH chains, i.e., the KH ladder, is getting more similar to that of the 2D honeycomb-lattice KH model. Only recognizable differences are the followings:\\
(i) The Ne\'el phase is replaced by rung-singlet phase. The rung-singlet gap decreases with increasing the number of KH chains and goes to zero in the honeycomb KH limit. This is essentially the same as the relation between $n$-leg Heisenberg ladder and 2D Heisenberg model.\\
(ii) The KSL phases in the KH ladder is wider than those in the honeycomb-lattice KH model because the quantum fluctuations are stronger due to the low dimensionality.

Finally, to get further insights into the topological properties of our system \eqref{ham}, we investigate the entanglement spectrum~\cite{li08}. Using Schmidt decomposition, the ground state can be expressed as
\begin{align}
|\psi=\sum_i e^{-\xi_i/2}|\phi_i^{\rm A}\rangle \otimes |\phi_i^{\rm B}\rangle,
\label{xi}
\end{align}
where the states $|\phi_i^S\rangle$ correspond to an orthonormal basis for the subsystem $S$ (either A or B).
We study a periodic ladder with $L\times2=32\times2$ sites and divide it into isometric subdomains A and B with $\frac{L}{2}\times2$ sites. In our calculations, the ES $\{\xi_i\}$ is simply obtained as $\xi_i=-\log \lambda_i$, where $\{\lambda_i\}$ are the eigenvalues of the reduced density matrices after the bipartite splitting. The low-lying entanglement spectrum levels are plotted as function of $\phi$ in Fig.~\ref{ES}. We find that the lowest level has no degeneracy in the magnetic LRO phases, which are topologically trivial. In the KSL phases, the lowest level has two-fold degeneracy and of the higher levels have high degrees of degeneracy. These are consistent with the ground-state phase diagram.

\section{Low-lying excitations}

\begin{figure}
  \includegraphics[width=0.9\columnwidth]{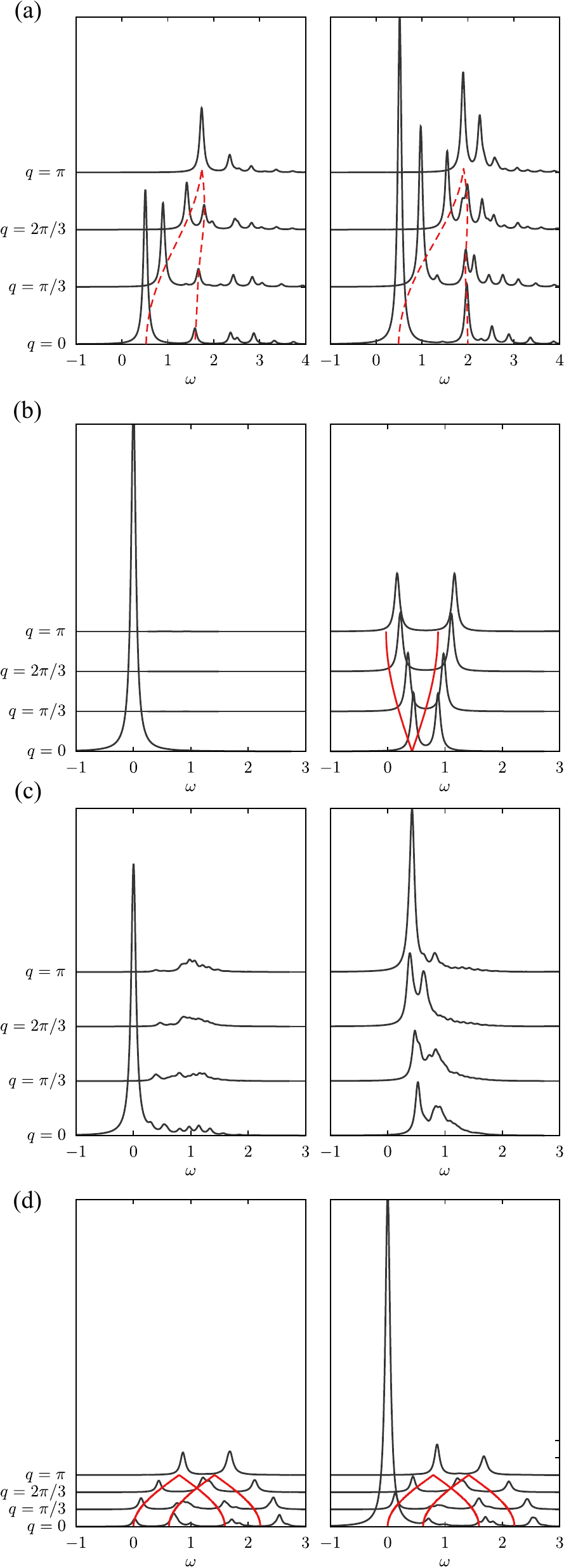}
  \caption{Dynamical structure factors calculated with a $12\times2$ ladder with periodic boundary conditions in the (a) rung-singlet ($\phi=0.2\pi$), (b) stripy ($\phi=1.64\pi$), (c) zigzag ($\phi=0.6\pi$), and (d) FM-$xy$ ($\phi=0.9\pi$) phases. The left and right panels show $S^z(q,\omega$ and $S^-(q,\omega)$, respectively. The red dotted lines are guide to the eye and red solid lines are spin-triplet dispersion obtained by the spin-wave theory.
  }
  \label{dsf-1}
\end{figure}

\begin{figure}
  \includegraphics[width=0.9\columnwidth]{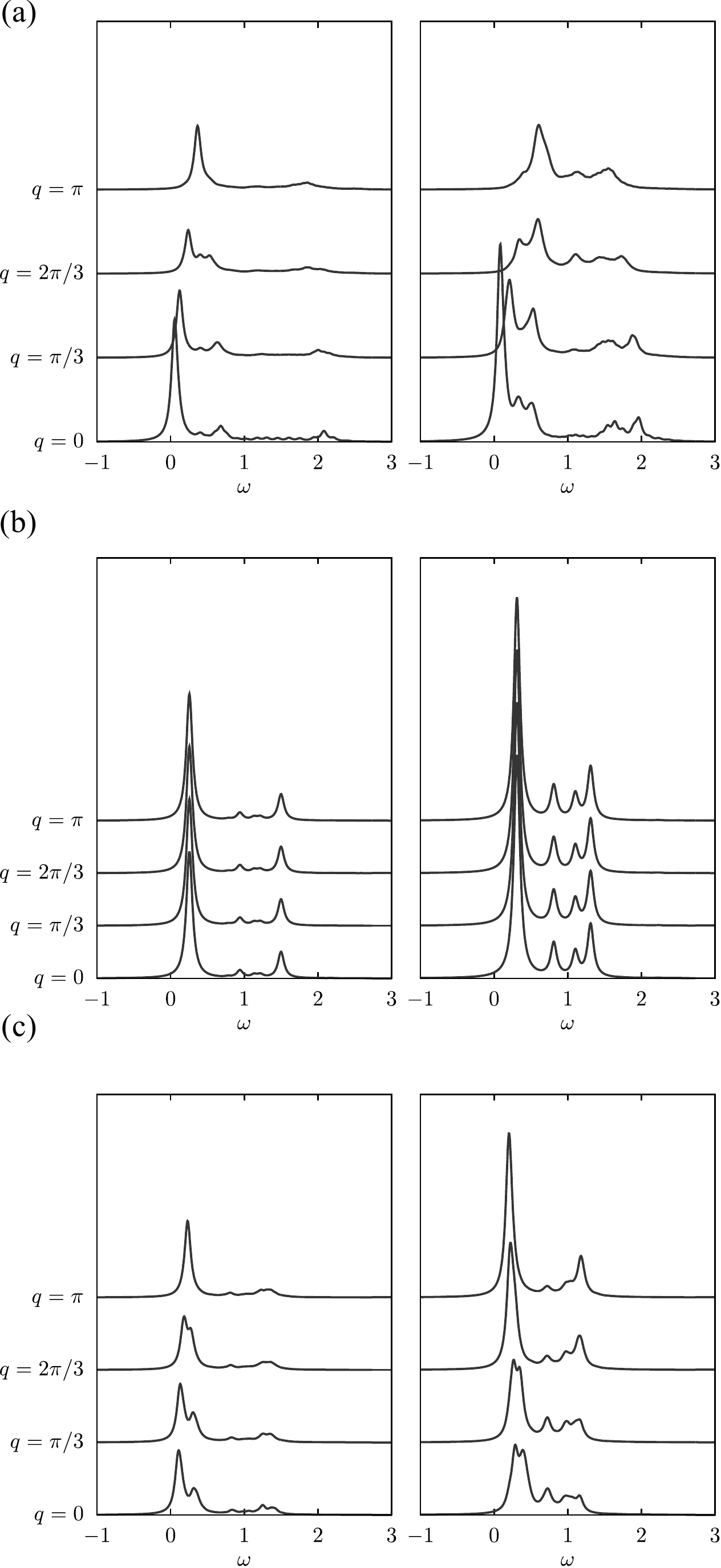}
  \caption{Dynamical structure factors in the FM KSL phase, calculated with a $12\times2$ ladder under periodic boundary conditions for (a) $\phi=1.4\pi$, (b) $\phi=1.5\pi$, and (c) $\phi=1.54\pi$. The left and right panels show $S^z(q,\omega$ and $S^-(q,\omega)$, respectively. The results for the AFM Kitaev point $\phi=0.5\pi$ are exactly the same as in (b). }
  \label{dsf-FMK}
\end{figure}

In this section, we study the low-lying excitations of the KH ladder by calculating the dynamical spin structure factor. We compute both $S^z(q, \omega)$ and $S^-(q,\omega)$ for each of the LRO phase. For the FM KSL state we compute them at three different $\phi$ values to study the effect of the Heisenberg interaction on the dispersion. The calculations were done using ED and a ladder of size $L=12\times2$ with periodic boundary conditions.

\subsection{Rung-singlet phase}

Fig.~\ref{dsf-1}(a) shows the dynamical structure factors for the rung-singlet state at $\phi=0.2\pi$  ($J\sim0.81,K\sim0.59$). The largest peak appears in $S^-(q=0,\omega\sim0.6)$ reflecting the dominant AFM fluctuations along the leg. The value of $\omega\sim0.6$ corresponds to the excitation gap $\Delta$ estimated above. The intensities in $S^-(q,\omega)$ are larger than those in $S^z(q,\omega)$ due to the easy-plane $xy$ anisotropy. As indicated in dotted line, the spin-triplet dispersion $\omega(q)$ of $S^z(q,\omega)$ can be well explained by that of the 2-leg Heisenberg ladder with the ratio between rung and leg interactions
$\frac{\rm rung}{\rm leg}\approx\frac{J+K}{J}\sim1.7$ (Ref.~\onlinecite{Barnes93}). The spin-triplet dispersion of $S^-(q,\omega)$ is similar in shape but it splits with the width $\sim \pm\frac{K}{2}$ at $q=\pi$. This splitting of spin-triplet dispersion is a general feature in system including the sign-alternating double-spin-flip term~\cite{Agrapidis18}. The width of spin-triplet dispersion in $S^z(q,\omega)$ and $S^-(q,\omega)$ roughly scales to $J$ and $J+\frac{K}{2}$, respectively.

\subsection{Stripy phase}

Fig.~\ref{dsf-1}(b) shows the dynamical structure factors for the stripy state at $\phi=1.64\pi$ ($J\sim0.43,K\sim-0.9$) where the single leg can be basically regarded as an easy-axis AFM XXZ Heisenberg chain. In $S^z(q,\omega)$ the largest peak appears at $(q,\omega)=(0,0)$ due to the Ne\'el ordering along the leg. Very few weights in the other momenta prove the validity of Eq.~\eqref{wf_stripy} with almost perfect alignment of spins parallel or antiparallel to $z$-axis and very weak quantum fluctuations. Whereas, the spin-triplet dispersion of $S^-(q,\omega)$ is basically explained by a single magnon dispersion. Thus, the spectral weight is almost uniform for all $q$ values, and the dispersion is obtained by spin-wave theory as
\begin{align}
\omega(q)=J\pm\frac{K}{2}\sin\frac{q}{2}.
\label{sqw_stripy}
\end{align}
The good agreement can be seen in Fig.~\ref{dsf-1}(b). Since the stripy order parameter drops on both phase boundaries, Eq.~\eqref{sqw_stripy} would give at least qualitatively a good approximation for the spin-triplet dispersion in the whole stripy phase.

\subsection{Zigzag phase}

Fig.~\ref{dsf-1}(c) shows the dynamical structure factors for the zigzag state at $\phi=0.6\pi (J\sim-0.31,K\sim0.95)$ where each leg is ferromagnetically ordered. The system can be understood as two FM chains coupled by the Ising-like AFM coupling. The largest peak in $S^z(q=0,\omega\sim0)$ reflects the FM ordering along the leg. The weights in the other momenta are small since the spins are mostly aligned along the $z$-axis; however, they seem to be larger than those for the stripy state. This implies that the zigzag ordering is more fragile than the stripy ordering. In $S^-(q,\omega)$ a largest and lowest-lying peak appears at $q=\pi$, indicating a four-site periodicity along the leg. The shape of the dispersion is similar to that of the staggered-$xy$ ordered state in the 1D KH model. Nevertheless, the gapped peak position $\omega\sim0.42$ clearly suggests no ordering on the $xy$-plane. The intensities in $S^-(q,\omega)$ are larger than those in $S^z(q,\omega)$ due to the easy-plane $xy$ anisotropy.

\subsection{FM states}

Fig.~\ref{dsf-1}(d) shows the dynamical structure factors for the FM-$xy$ state at $\phi=0.9\pi (J\sim-0.95,K\sim0.30)$. The largest peak in $S^-(q=0,\omega\sim0)$ confirms that the spins lie mostly on the $xy$-plane. The other features are very similar between $S^z(q,\omega)$ and $S^-(q,\omega)$. Both of them have the same excitation dispersion as
\begin{align}
\omega_1(q)=-\frac{2J+K}{2}\left(1\pm\cos\frac{q}{2}\right),
\label{sqw_FM1}
\end{align}
and
\begin{align}
\omega_2(q)=-\frac{2J+K}{2}\left(1\pm\cos\frac{q}{2}\right)+2|K|.
\label{sqw_FM2}
\end{align}
The splitting between $\omega_1(q)$ and $\omega_2(q)$ becomes zero in the isotropic SU(2) summetric point at $q=\pi$ and it is roughly proportional to $|K|$.

\subsection{Kitaev spin liquid}

In Fig.~\ref{dsf-FMK} we show the dynamical structure factors around the FM Kitaev point ($\phi=\frac{3}{2}\pi$). At the FM Kitaev point, both $S^z(q,\omega)$ and $ S^-(q,\omega)$ show no dependence on $q$. This dispersionless feature is a natural consequence of no spin-spin correlations except the nearest-neighbor ones. The distance between the lower and upper bound of the continuum in $S^-(q,\omega)$ is of the order of $|K|$, relating to the spinon propagation along the leg. Note that the spectra at the AFM Kitaev point are exactly the same as those at the FM Kitaev point. Let us then see the effect of the Heisenberg term on the spectra. Fig.~\ref{dsf-FMK}(a) and (c) show the dynamical structure factors at $\phi=1.4\pi$ and $\phi=1.54\pi$, respectively. Although they are almost equally close to the boundary to the neighboring phase, the spectra are apparently quite different: At $\phi=1.54$ it mostly keeps the spectral features at the Kitaev point except that the main peak splits into two peaks with a small interval $\sim J$; while at $\phi=1.4\pi$ the dispersionless feature is completely collapsed and its lower bound looks rather similar to the spin-triplet dispersion of the FM-$xy$ state. It may be related to the fact that the expectation value of the plaquette operator deviates faster from the pure KSL value ($\mathcal{O}_{\rm plaquette}=1$) at $\phi<\frac{3}{2}\pi$ ($J<0$) than at $\phi<\frac{3}{2}\pi$ ($J>0$) with leaving from the FM Kitaev point.

%\subsection{Entanglement spectra}

\section{Conclusion}

We studied the ground state and low-lying excitations of the KH model on a ladder using the DMRG and Lanczos ED methods. Based on the results of several order parameters, excitation gap, and entanglement spectra, we determined the ground-state phase diagram as a function of the ratio between Kitaev and Heisenberg interactions. It is very rich and includes four magnetically ordered phases such as rung-singlet, stripy, zigzag, FM, and two spin liquid phases. The phase diagram is strikingly similar to that of the KH model on a honeycomb lattice. Distinct differences
are only the presence of a rung-singlet phase instead of the N\'eel state and a few times wider ranges of two spin liquid phases. These differences can be understood by the dimensionality: (i) Since the quantum fluctuations are typically stronger in a lower dimensional system, it is more difficult to stabilize LRO state in the ladder than in a 2D system. (ii) Though the 2-leg KH ladder has a finite excitation gap in the rung-singlet phase around $\phi=0$ due to strong cluster anisotropy, the gap decreases with increasing the number of legs and becomes zero in the limit of 2D honeycomb-lattice KH model. We also calculated the dynamical spin structure factor using the Lanczos ED method. Interestingly, most of the spectral features in the KH ladder can be explained by considering those of the 1D KH model.

Note added --- During the preparation of this manuscript, we became aware of Ref.~\onlinecite{Catuneanu18}. Their phase diagram agrees very well with ours.

\bibliography{KHladder}

\end{document}